\documentclass{emulateapj}
\newcommand{\etal}{et al.\ }
\newcommand{\etalb}{et al.}
\newcommand{\be}{\begin{equation}}
\newcommand{\ba}{\begin{eqnarray}}
\newcommand{\ee}{\end{equation}}
\newcommand{\ea}{\end{eqnarray}}
\newcommand{\del}{\delta}

\usepackage{apjfonts}

\begin{document}
\title{Unusually Large Fluctuations in the Statistics \\ of Galaxy
Formation at High Redshift}

\author{Rennan Barkana}
\affil{School of Physics and Astronomy, The Raymond and Beverly Sackler
Faculty of Exact Sciences, \\
Tel Aviv University, Tel Aviv 69978, ISRAEL; barkana@wise.tau.ac.il}

\author{Abraham Loeb} 
\affil{Astronomy Department, Harvard University, 60 
Garden Street, Cambridge, MA 02138; aloeb@cfa.harvard.edu}

\begin{abstract} 
We show that various milestones of high-redshift galaxy formation,
such as the formation of the first stars or the complete reionization
of the intergalactic medium, occurred at different times in different
regions of the universe. The predicted spread in redshift, caused by
large-scale fluctuations in the number density of galaxies, is at
least an order of magnitude larger than previous expectations that
argued for a sharp end to reionization.  This cosmic scatter in the
abundance of galaxies introduces new features that affect the nature
of reionization and the expectations for future probes of
reionization, and may help explain the present properties of dwarf
galaxies in different environments. The predictions can be tested by
future numerical simulations and may be verified by upcoming
observations. Current simulations, limited to relatively small volumes
and periodic boundary conditions, largely omit cosmic scatter and its
consequences. In particular, they artificially produce a sudden end to
reionization, and they underestimate the number of galaxies by up to
an order of magnitude at redshift 20.
\end{abstract}

\keywords{galaxies: high-redshift, cosmology: theory, galaxies:
formation}

\section{Introduction}

Recent observations of the cosmic microwave background \citep{WMAP}
have confirmed the notion that the present large-scale structure in
the universe originated from small-amplitude density fluctuations at
early cosmic times. Due to the natural instability of gravity, regions
that were denser than average collapsed and formed bound halos, first
on small spatial scales and later on larger and larger scales. At each
snapshot of this cosmic evolution, the abundance of collapsed halos,
whose masses are dominated by cold dark matter, can be computed from
the initial conditions using numerical simulations and can be
understood using approximate analytic models \citep{ps74, bond91}. The
common understanding of galaxy formation is based on the notion that
the constituent stars formed out of the gas that cooled and
subsequently condensed to high densities in the cores of some of these
halos \citep{wr78}.

The standard analytic model for the abundance of halos \citep{ps74,
bond91} considers the small density fluctuations at some early,
initial time, and attempts to predict the number of halos that will
form at some later time corresponding to a redshift $z$. First, the
fluctuations are extrapolated to the present time using the growth
rate of linear fluctuations, and then the average density is computed
in spheres of various sizes. Whenever the overdensity (i.e., the
density perturbation in units of the cosmic mean density) in a sphere
rises above a critical threshold $\delta_c(z)$, the corresponding
region is assumed to have collapsed by redshift $z$, forming a halo
out of all the mass that had been included in the initial spherical
region. In analyzing the statistics of such regions, the model
separates the contribution of large-scale modes from that of
small-scale density fluctuations. It predicts that galactic halos will
form earlier in regions that are overdense on large scales \citep{k84,
b86, ck89, mw96}, since these regions already start out from an
enhanced level of density, and small-scale modes need only supply the
remaining perturbation necessary to reach $\delta_c(z)$. On the other
hand, large-scale voids should contain a reduced number of halos at
high redshift. In this way, the analytic model describes the
clustering of massive halos.

As gas falls into a dark matter halo, it can fragment into stars only
if its virial temperature is above $10^4$K for cooling mediated by
atomic transitions [or $\sim 500$ K for molecular ${\rm H}_2$ cooling;
see, e.g., Figure~12 in \citet{review}]. The abundance of dark matter
halos with a virial temperature above this cooling threshold declines
sharply with increasing redshift due to the exponential cutoff in the
abundance of massive halos at early cosmic times. Consequently, a
small change in the collapse threshold of these rare halos, due to
mild inhomogeneities on much larger spatial scales, can change the
abundance of such halos dramatically. The modulation of galaxy
formation by long wavelength modes of density fluctuations is
therefore amplified considerably at high redshift. In this paper we
show that this results in major new predictions for high-redshift
observations. The implications are particularly significant for cosmic
reionization and all observational probes of this epoch.

This paper is organized as follows. In \S~2 we quantify the scatter in
the statistics of galaxy formation produced by this amplification
effect. We first explain in \S~2.1 the basic physical ideas and
implications using the well-established extended Press-Schechter
model. We then present in \S~2.2 a simple idea that yields a much more
accurate model that fits an array of previous simulations at low
redshift. We demonstrate the qualitative correctness of our basic
assumptions as well as the quantitative accuracy of our model by
matching results from recent simulations at high redshift. Since
high-redshift galaxies provide the UV photons that lead to the
reionization of the intergalactic medium (hereafter IGM), a large
scatter is also expected in the reionization redshift within different
regions in the universe. We consider this scatter and the modified
character of reionization in \S~3.1, and show in \S~3.2 that existing
numerical simulations do not include fluctuations on sufficiently
large scales at high redshift. In \S~3.3 we discuss the observational
implications of the large cosmic scatter expected at high redshift.
Finally, we summarize our main results in \S~4.

\section{Halo Mass Function in Different Environments} 

\subsection{Basic Model: Amplification of Density Fluctuations}

Galaxies at high redshift are believed to form in all halos above some
minimum mass $M_{\rm min}$ that depends on the efficiency of atomic
and molecular transitions that cool the gas within each halo. This
makes useful the standard quantity of the collapse fraction $F_{\rm
col}(M_{\rm min})$, which is the fraction of mass in a given volume
that is contained in halos of individual mass $M_{\rm min}$ or
greater. If we set $M_{\rm min}$ to be the minimum halo mass in which
efficient cooling processes are triggered, then $F_{\rm col}(M_{\rm
min})$ is the fraction of all the baryons in the universe that lie in
galaxies. In a large-scale region of comoving radius $R$ with a mean
overdensity $\bar{\delta}_R$, the standard result is \citep{bond91}
\begin{equation} F_{\rm col}(M_{\rm min})={\rm erfc}\left[
\frac{\delta_c(z)- \bar{\delta}_R} {\sqrt{2 \left[S(R_{\rm min}) -
S(R) \right]}} \right]\ , \label{eq:Fcol} \end{equation} where $S(R)$
is the variance of fluctuations in spheres of radius $R$, and
$S(R_{\rm min})$ is the variance in spheres of radius $R_{\rm min}$
corresponding to the region at the initial time that contained a mass
$M_{\rm min}$. In particular, the cosmic mean value of the collapse
fraction is obtained in the limit of $R \rightarrow \infty$ by setting
$\bar{\delta}_R$ and $S(R)$ to zero in this expression. Throughout
this section our results assume this standard model, known as the
extended Press-Schechter model, which we apply to a universe with
cosmological parameters matching the latest observations
[specifically, the running index model of \citet{WMAP}]. Whenever we
consider a cubic region, we estimate its halo abundance by applying
the model to a spherical region of equal volume. Note also that we
consistently quote values of comoving distance, which equals physical
distance times a factor of $(1+z)$.

Our results are based on a simple idea. At high redshift, galactic
halos are rare and correspond to high peaks in the Gaussian
probability distribution of initial fluctuations. A modest change in
the overall density of a large region modulates the threshold for high
peaks in the Gaussian density field, so that the number of galaxies is
exponentially sensitive to this modulation. This amplification of
large-scale modes is responsible for the large statistical
fluctuations that we find.

In numerical simulations, periodic boundary conditions are usually
assumed, and this forces the mean density of the box to equal the
cosmic mean density. The abundance of halos as a function of mass is
then biased in such a box (see Figure~1), since a similar region in
the real universe will have a distribution of different overdensities
$\bar{\delta}_R$.  At high redshift, when galaxies correspond to high
peaks, they are mostly found in regions with an enhanced large-scale
density. In a periodic box, therefore, the total number of galaxies is
artificially reduced, and the relative abundance of galactic halos
with different masses is artificially tilted in favor of lower-mass
halos. We illustrate our results for two sets of parameters, one
corresponding to the first galaxies and early reionization ($z=20$)
and the other to the current horizon in observations of galaxies and
late reionization ($z=7$). We consider a resolution equal to that of
state-of-the-art cosmological simulations that include gravity and gas
hydrodynamics. Specifically, we assume that the total number of dark
matter particles in the simulation is $N = 324^3$, and that the
smallest halo that can form a galaxy must be resolved into 500
particles; \citet{converge} showed that this resolution is necessary
in order to determine the star formation rate in an individual halo
reliably to within a factor of two. Therefore, if we assume that halos
that cool via molecular hydrogen must be resolved at $z=20$ (so that
$M_{\rm min}=7 \times 10^5 M_{\odot}$), and only those that cool via
atomic transitions must be resolved at $z=7$ (so that $M_{\rm
min}=10^8 M_{\odot}$), then the maximum box sizes that can currently
be simulated are $l_{\rm box}=1$ Mpc and $l_{\rm box}=6$ Mpc at these
two redshifts, respectively.

\begin{figure} 
\plotone{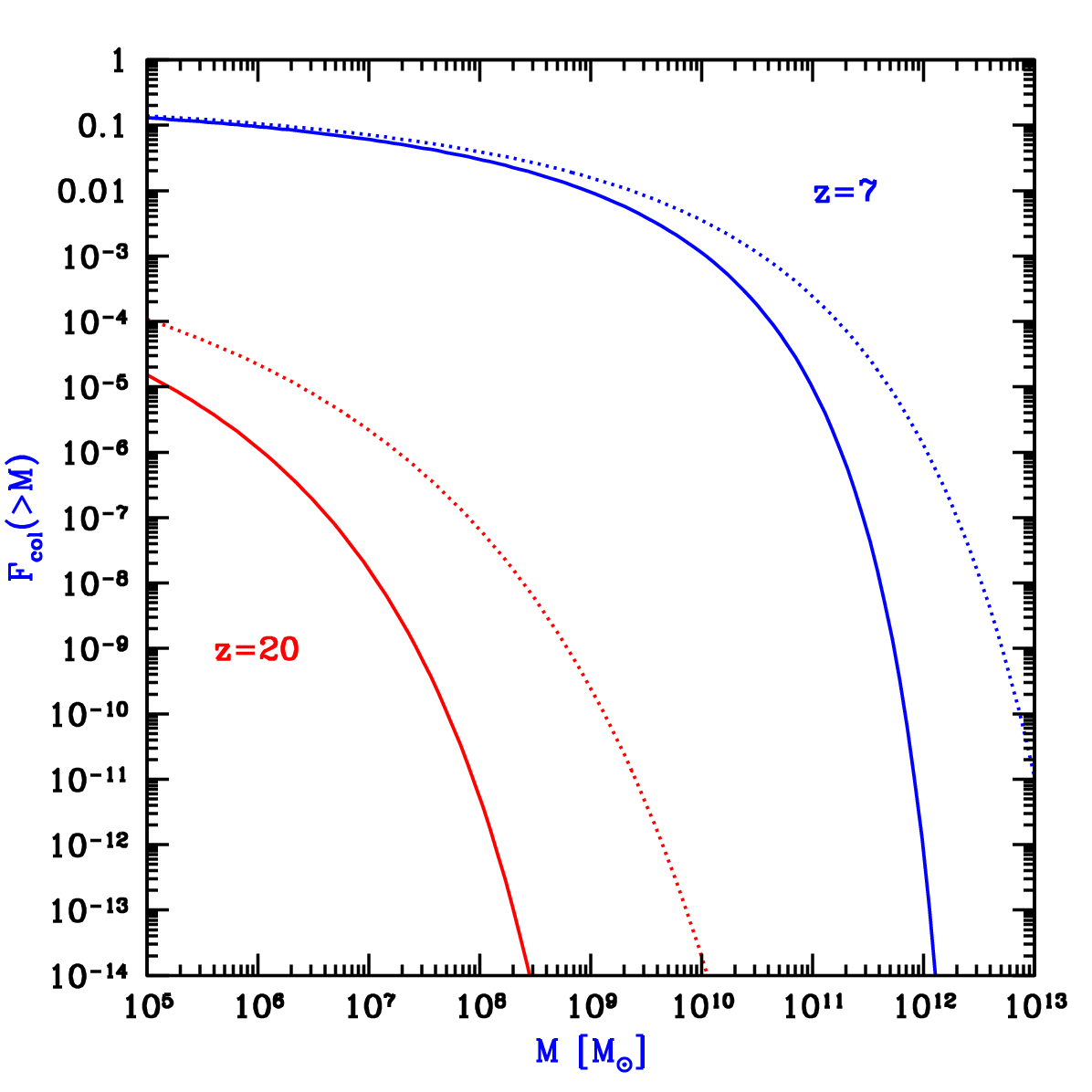} 
\caption{Bias in the halo mass distribution in simulations. We show
the amount of mass contained in all halos of individual mass $M_{\rm
min}$ or greater, expressed as a fraction of the total mass in a given
volume. This cumulative fraction $F_{\rm col}(M_{\rm min})$ is shown
as a function of the minimum halo mass $M_{\rm min}$. We consider two
cases of redshift and simulation box size, namely $z=7$, $l_{\rm
box}=6$ Mpc (upper curves), and $z=20$, $l_{\rm box}=1$ Mpc (lower
curves). At each redshift, we compare the true average distribution in
the universe (dotted curve) to the biased distribution (solid curve)
that would be measured in a simulation box with periodic boundary
conditions (for which $\bar{\delta}_R$ is artificially set to zero).}
\end{figure}

At each redshift we only consider cubic boxes large enough so that the
probability of forming a halo on the scale of the entire box is
negligible. In this case, $\bar{\delta}_R$ is Gaussian distributed
with zero mean and variance $S(R)$, since the no-halo condition
$\sqrt{S(R)} \ll \del_c(z)$ implies that at redshift $z$ the
perturbation on the scale $R$ is still in the linear regime. We can
then calculate the probability distribution of collapse fractions in a
box of a given size (see Figure~2). This distribution corresponds to a
real variation in the fraction of gas in galaxies within different
regions of the universe at a given time. In a numerical simulation,
the assumption of periodic boundary conditions eliminates the
large-scale modes that cause this cosmic scatter. Note that Poisson
fluctuations in the number of halos within the box would only add to
the scatter, although the variations we have calculated are typically
the dominant factor. For instance, in our two standard examples, the
mean expected number of halos in the box is 3 at $z=20$ and 900 at
$z=7$, resulting in Poisson fluctuations of a factor of about 2 and
1.03, respectively, compared to the clustering-induced scatter of a
factor of about 16 and 2 in these two cases.

\begin{figure}
\plotone{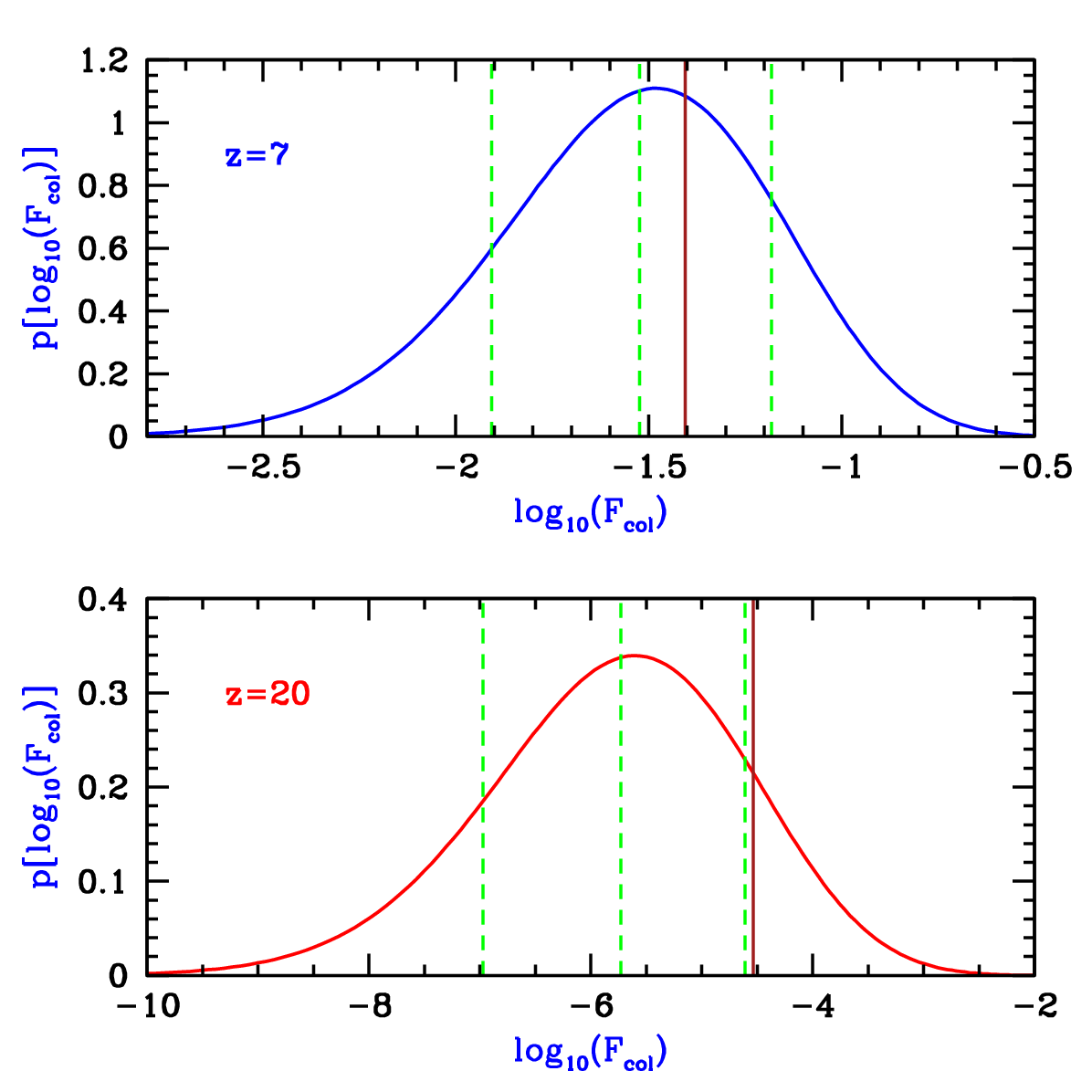}
\caption{Probability distribution within a small volume of the total
mass fraction in galactic halos. The normalized distribution of the
logarithm of this fraction $F_{\rm col}(M_{\rm min})$ is shown for two
cases: $z=7$, $l_{\rm box}=6$ Mpc, $M_{\rm min}=10^8 M_{\odot}$ (upper
panel), and $z=20$, $l_{\rm box}=1$ Mpc, $M_{\rm min}=7 \times 10^5
M_{\odot}$ (bottom panel). In each case, the value in a periodic box
($\bar{\delta}_R=0$) is shown (central dashed vertical line), along
with the value that would be expected given a plus $1-\sigma$ (right
dashed line) or a minus $1-\sigma$ (left dashed line) fluctuation in
the mean density of the box. Also shown in each case is the mean value
of $F_{\rm col}(M_{\rm min})$ averaged over large cosmological volumes
(solid vertical line).}
\end{figure}

Within the extended Press-Schechter model, both the numerical bias and
the cosmic scatter can be simply described in terms of a shift in the
redshift (see Figure~3). In general, a region of radius $R$ with a
mean overdensity $\bar{\delta}_R$ will contain a different collapse
fraction than the cosmic mean value at a given redshift $z$. However,
at some wrong redshift $z + \Delta z$ this small region will contain
the cosmic mean collapse fraction at $z$. At high redshifts ($z > 3$),
this shift in redshift can be easily derived from eq.~1 to be
\begin{equation} \Delta z = \frac{\bar{\delta}_R}{\delta_0} - (1+z)
\times \left[ 1 - \sqrt{1 - \frac{S(R)} {S(R_{\rm min})}}\ \right]\ ,
\end{equation} where $\delta_0 \equiv \delta_c(z)/(1+z)$ is
approximately constant at high redshifts \citep{p80}, and equals 1.28
for our assumed cosmological parameters. Thus, in our two standard
examples, the bias is -2.6 at $z=20$ and -0.4 at $z=7$, and the
one-sided $1-\sigma$ scatter is 2.4 at $z=20$ and 1.2 at $z=7$.

\begin{figure}
\plotone{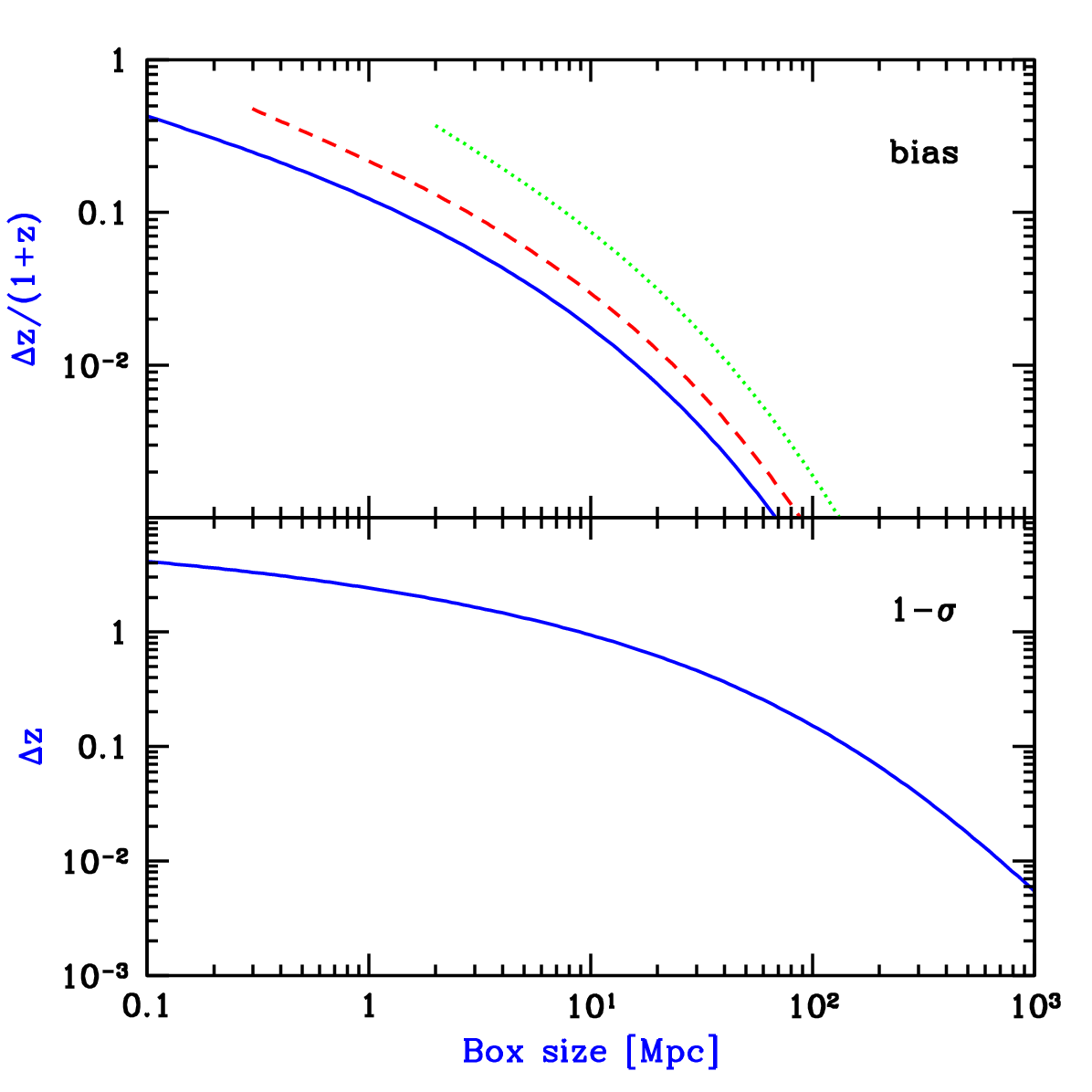}
\caption{Cosmic scatter and numerical bias, expressed as the change in
redshift needed to get the correct cosmic mean of the collapse
fraction. We show the $1-\sigma$ scatter (about the biased value) in
the redshift of reionization, or any other phenomenon that depends on
the mass fraction in galaxies (bottom panel), as well as the redshift
bias [expressed as a fraction of $(1+z)$] in periodic simulation boxes
(upper panel). The bias is shown for $M_{\rm min}=7 \times 10^5
M_{\odot}$ (solid curve), $M_{\rm min}=10^8 M_{\odot}$ (dashed curve),
and $M_{\rm min}=3 \times 10^{10} M_{\odot}$ (dotted curve). The bias
is always negative, and we show its absolute value. When expressed as
a shift in redshift, the scatter is independent of $M_{\rm min}$.}
\end{figure}

\subsection{Improved Model: Matching Numerical Simulations}

\label{massfn}

In this subsection we develop an improved model that fits the results
of numerical simulations more accurately. The model constructs the
halo mass distribution (or mass function); cumulative quantities such
as the collapse fraction or the total number of galaxies can then be
determined from it via integration. We first define $f(\del_c(z),S)\,
dS$ to be the mass fraction contained at $z$ within halos with mass in
the range corresponding to $S$ to $S+d S$. The halo abundance is then
\be \frac{dn}{dM} = \frac{\bar{\rho}_0}{M} \left|\frac{d S}{d M}
\right| f(\del_c(z),S)\ , \label{eq:abundance} \ee where $dn$ is the
comoving number density of halos with masses in the range $M$ to
$M+dM$. In the model of \citet{ps74}, \be f_{\rm PS}(\del_c(z),S) =
\frac{1} {\sqrt{2 \pi}} \frac{\nu }{S} \exp\left[-\frac{\nu^2}{2}
\right]\ , \label{eq:PS} \ee where $\nu=\del_c(z)/\sqrt{S}$ is the
number of standard deviations that the critical collapse overdensity
represents on the mass scale $M$ corresponding to the variance $S$.

However, the Press-Schechter mass function fits numerical simulations
only roughly, and in particular it substantially underestimates the
abundance of the rare halos that host galaxies at high redshift. The
halo mass function of \citet[][see also \citealt{shethmot}]{shetht99}
adds two free parameters that allow it to fit numerical simulations
much more accurately \citep{jenkins}. We note that these simulations
followed very large volumes at low redshift, so that cosmic scatter
did not compromise their accuracy. The matching mass function is given
by \be f_{\rm ST}(\del_c(z),S) = A' \frac{\nu }{S} \sqrt{\frac{a'} {2
\pi}} \left[ 1+\frac{1}{(a' \nu^2)^{q'}} \right] \exp\left[-\frac{a'
\nu^2}{2} \right]\ , \label{eq:ST} \ee with best-fit parameters
\citep{shetht02} $a'=0.75$ and $q'=0.3$, and where normalization to
unity is ensured by taking $A'=0.322$.

In order to calculate cosmic scatter we must determine the biased halo
mass function in a given volume at a given mean density. Within the
extended Press-Schechter model \citep{bond91}, the halo mass
distribution in a region of comoving radius $R$ with a mean
overdensity $\bar{\delta}_R$ is given by \be f_{\rm
bias-PS}(\del_c(z), \bar{\delta}_R,R,S)=f_{\rm PS}(\del_c(z)-
\bar{\delta}_R,S-S(R))\ \label{eq:ePS}. \ee The corresponding collapse
fraction in this case is given simply by eq.~(\ref{eq:Fcol}). Despite
the relatively low accuracy of the Press-Schechter mass function, the
{\it relative change} is predicted rather accurately by the extended
Press-Schechter model. In other words, the prediction for the halo
mass function in a given volume compared to the cosmic mean mass
function provides a good fit to numerical simulations over a wide
range of parameters \citep{mw96,casas02, shetht02}.

\begin{figure}[htbp] 
\plotone{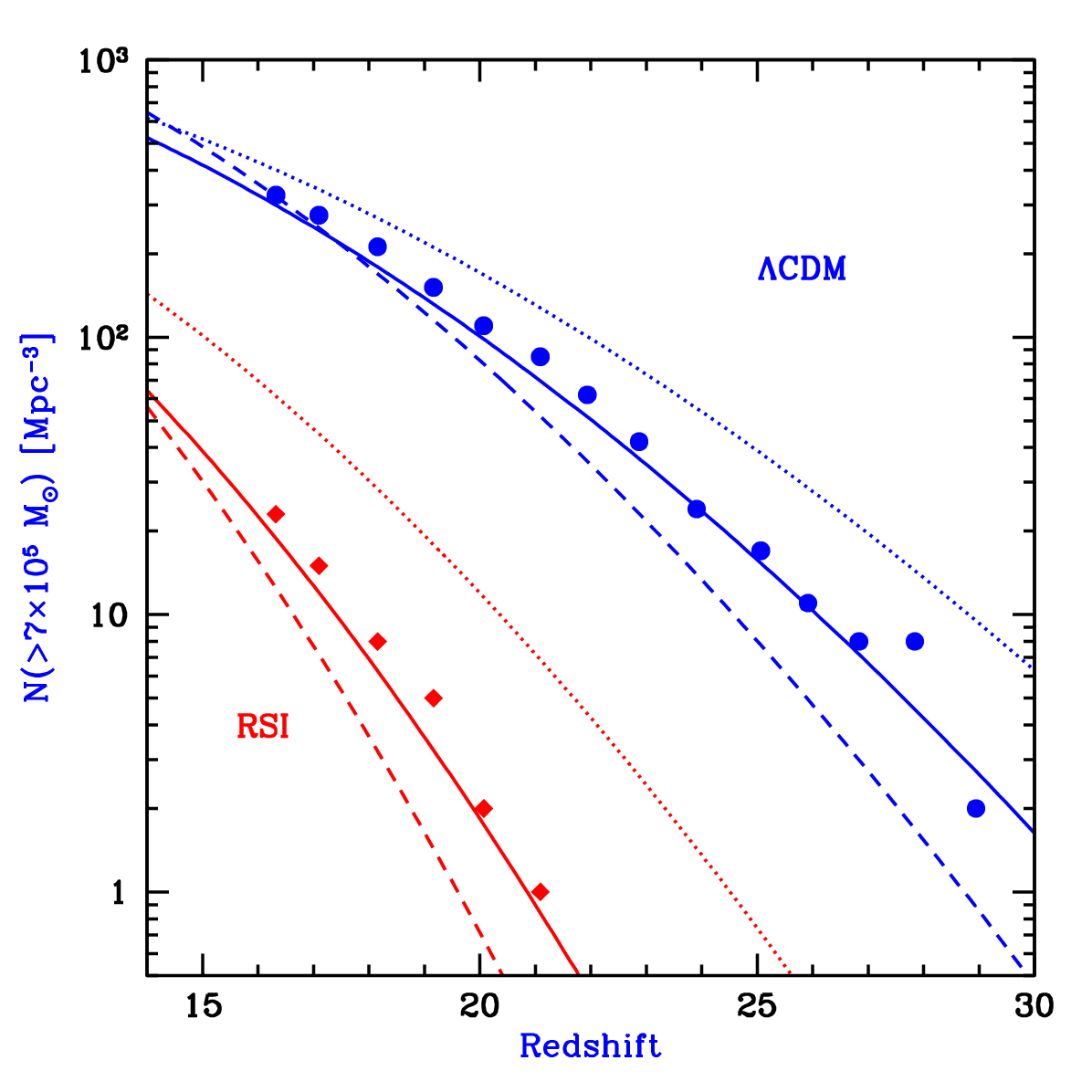} 
\caption{Halo mass function at high redshift in a 1 Mpc box at the
cosmic mean density. Our hybrid model prediction (solid lines) is
compared with the number of halos above mass $7 \times 10^5 M_{\odot}$
measured in the simulations of \citet[][data points are taken from
their Figure~5]{yoshida}. The cosmic mean of the halo mass function
(dotted lines) deviates significantly from the simulated values, since
the periodic boundary conditions within the finite simulation box
artificially set the amplitude of large-scale modes to zero. Our
hybrid model starts with the Sheth-Tormen mass function and applies a
correction based on the extended Press-Schechter model; in doing so,
it provides a better fit to numerical simulations than the pure
extended Press-Schechter model (dashed lines) used in the previous
figures. We consider two sets of cosmological parameters, the
scale-invariant $\Lambda$CDM model of \citet{yoshida} (upper curves),
and their running scalar index (RSI) model (lower curves).}
\label{fig-Lars} 
\end{figure}

For our improved model we adopt a hybrid approach that combines
various previous models with each applied where it has been found to
closely match numerical simulations. We obtain the halo mass function
within a restricted volume by starting with the Sheth-Tormen formula
for the cosmic mean mass function, and then adjusting it with a
relative correction based on the extended Press-Schechter model. In
other words, we set \ba & & f_{\rm bias}(\del_c(z),\bar{\delta}_R,R,S)
= \nonumber \\ & & f_{\rm ST}(\del_c(z),S)\ \times \left[ \frac{f_{\rm
PS} (\del_c(z)-\bar{\delta}_R,S-S(R))} {f_{\rm PS}(\del_c(z),S)}
\right]\ . \label{eq:bias} \ea As noted, this model is based on fits
to simulations at low redshifts, but we can check it at high redshifts
as well. Figure~\ref{fig-Lars} shows the number of galactic halos at
$z \sim 15-30$ in two numerical simulations run by \citet{yoshida},
and our predictions given the cosmological input parameters assumed by
each simulation. The close fit to the simulated data (with no
additional free parameters) suggests that our hybrid model (solid
lines) improves on the extended Press-Schechter model (dashed lines),
and can be used to calculate accurately the cosmic scatter in the
number of galaxies at both high and low redshifts. The simulated data
significantly deviate from the expected cosmic mean
[eq.~(\ref{eq:ST}), shown by the dotted line], due to the artificial
suppression of large-scale modes outside the simulated box. We note
that \citet{yoshida} mentioned that the lack of large-scale modes
might produce a systematically low halo abundance, particularly in the
RSI model, but they did not quantify this effect.

As an additional example, we consider the highest-resolution first
star simulation \citep{abn02}, which used $l_{\rm box}=128$ kpc and
$M_{\rm min}=7 \times 10^5 M_{\odot}$.  The first star forms within
the simulated volume when the first halo of mass $M_{\rm min}$ or
larger collapses within the box. To compare with the simulation, we
predict the redshift at which the probability of finding at least one
halo within the box equals $50\%$, accounting for Poisson
fluctuations. We find that if the simulation formed a population of
halos corresponding to the correct cosmic average [as given by
eq.~(\ref{eq:ST})], then the first star should have formed already at
$z=24.0$. The first star actually formed in the simulation box only at
$z=18.2$ \citep{abn02}. Using eq.~(\ref{eq:bias}) we can account for
the loss of large-scale modes beyond the periodic box, and predict a
first star at $z=17.8$, a close match given the large Poisson
fluctuations introduced by considering a single galaxy within the box.

The artificial bias in periodic simulation boxes can also be seen in
the results of extensive numerical convergence tests carried out by
\citet{converge}. They presented a large array of numerical
simulations of galaxy formation run in periodic boxes over a wide
range of box size, mass resolution, and redshift. In particular, we
can identify several pairs of simulations where the simulations in
each pair have the same mass resolution but different box sizes; this
allows us to separate the effect of large-scale numerical bias from
the effect of having poorly-resolved individual halos. Specifically,
their simulations Z1 and R4 [see Table~1 in \citet{converge}] used the
same particle mass but R4 had a box length larger by a factor of 3.375
. The simulations Q1 and D4 are similarly related, as are Q2 and
D5. In each case, the smaller simulation substantially underestimated
the star formation rate at high redshift [see Figure~10 in
\citet{converge}], Z1 by a factor of 5 at $z=15$ compared to R4, Q1 by
a factor of 3 at $z=8$ compared to D4, and Q2 by a factor of 3 at
$z=10$ compared to D5.

We note that there have been previous attempts to develop a model for
the halo mass function in different environments, so that the model
would be consistent with the Sheth-Tormen mass function of
eq.~(\ref{eq:ST}) which accurately fits the cosmic mean mass function
measured in numerical simulations. In order to identify the specific
requirements for such a consistency, we first consider the analogous
case of the extended Press-Schechter model and its relation to the
Press-Schechter formula for the cosmic mean mass function. The
extended Press-Schechter model is consistent with the mean mass
function in the sense that $f_{\rm bias-PS}$ evaluated for an infinite
box (i.e., in the limit where $\bar{\delta}_R$ and $S(R)$ both vanish)
yields the Press-Schechter mass function: \be f_{\rm bias-PS}
(\del_c(z), 0,\infty,S)=f_{\rm PS}(\del_c(z),S)\ . \label{eq:cons} \ee
This condition does not suffice, however, since there is an additional
self-consistency test that any viable model must satisfy. Consider any
fixed scale $R$. Suppose we consider a very large number $N$ of
spheres of radius $R$ within the universe. The mean density
$\bar{\delta}_R$ in each sphere is determined according to a
probability distribution $p(\bar{\delta}_R)$; we assume that $R$ is
large enough so that the probability of forming a halo out of all the
mass on the scale $R$ is negligible, and so the distribution is a
Gaussian with zero mean and variance $S(R)$ (see also \S 2.1). The
number of galaxies in each sphere is given in the extended
Press-Schechter model by eq.~(\ref{eq:ePS}). As $N \rightarrow
\infty$, the halo mass function averaged over all these spheres must
approach the cosmic mean value, and it must also approach the
ensemble-averaged mass function, where the averaging is performed over
the probability distribution of $\bar{\delta}_R$. This yields the
following self-consistency requirement, which is indeed satisfied by
the extended Press-Schechter model: \ba & & \int f_{\rm
bias-PS}(\del_c(z), \bar{\delta}_R,R,S)\, p(\bar{\delta}_R)\,
d\bar{\delta}_R = \nonumber \\ & & f_{\rm bias-PS}(\del_c(z),
0,\infty,S)\ .  \label{eq:scons} \ea

Now we again consider attempts to construct an improved model that is
consistent with the Sheth-Tormen mass function. Such a model must
satisfy eq.~(\ref{eq:cons}) (except with $f_{\rm ST}$ on the
right-hand side), and it must also satisfy eq.~(\ref{eq:scons}) in
order to be self-consistent. The latter equation must be satisfied
separately for every scale $R$ large enough to avoid collapsing [i.e.,
that satisfies $\sqrt{S(R)} \ll \del_c(z)$]. Previous proposed models
\citep{shetht02, gottl03} satisfied simple consistency but not the
self-consistency test. Our hybrid model of eq.~(\ref{eq:bias})
satisfies both requirements (with respect to the Sheth-Tormen mass
function), a result that follows immediately from the fact that the
extended Press-Schechter model also satisfies both requirements (with
respect to the Press-Schechter mass function). Thus, our hybrid model
is the first self-consistent model that is also consistent with the
Sheth-Tormen mass function, at least when fluctuations are considered
on large scales $R$ for which $\bar{\delta}_R$ is Gaussian
distributed. As demonstrated in this section, our model also matches
results from a wide array of numerical simulations
\citep{abn02, yoshida, mw96, casas02, jenkins}.

\section{Implications}

\subsection{The nature of reionization} 

The photons of the cosmic microwave background have traveled to us
mostly undisturbed after neutral atoms first formed in the universe at
the cosmic recombination epoch. Radiation from the first generation of
stars is thought to have reionized the hydrogen throughout the
universe, transforming the IGM back into a hot and highly-ionized
plasma.

The popular view developed in the literature \citep{aw72, fk94, sgb94,
hl97, g00, review} maintains that reionization ended with a fast,
simultaneous, overlap stage throughout the universe. This view has
been based on simple arguments and has been supported by numerical
simulations with small box sizes. The underlying idea was that the
ionized hydrogen (\ion{H}{2}) regions of individual sources began to
overlap when the typical size of each \ion{H}{2} bubble became
comparable to the distance between nearby sources. Since these two
length scales were comparable at the critical moment, there is only a
single timescale in the problem -- given by the growth rate of each
bubble -- and it determines the transition time between the initial
overlap of two or three nearby bubbles, to the final stage where
dozens or hundreds of individual sources overlap and produce large
ionized regions. Whenever two ionized bubbles were joined, each point
inside their common boundary became exposed to ionizing photons from
both sources, reducing the neutral hydrogen fraction and allowing
ionizing photons to travel farther before being absorbed. Thus, the
ionizing intensity inside \ion{H}{2} regions rose rapidly, allowing
those regions to expand into high-density gas that had previously
recombined fast enough to remain neutral when the ionizing intensity
had been low. Since each bubble coalescence accelerates the process,
it has been thought that the overlap phase has the character of a
phase transition and occurs rapidly. Indeed, the best simulations of
reionization to date \citep{g00} found that the average mean free path
of ionizing photons in the simulated volume rises by an order of
magnitude over a redshift interval $\Delta z = 0.05$ at $z=7$.

Our results substantially modify this commonly accepted picture for
the development of reionization. Overlap is still expected to occur
rapidly, but only in localized high-density regions, where the
ionizing intensity and the mean free path rise rapidly even while
other distant regions are still mostly neutral. In other words, the
size of the bubble of an individual source is about the same in
different regions (since most halos have masses just above $M_{\rm
min}$), but the typical distance between nearby sources varies widely
across the universe. The strong clustering of ionizing sources on
length scales as large as 30--100 Mpc introduces long timescales into
the reionization phase transition. The sharpness of overlap is
determined not by the growth rate of bubbles around individual
sources, but by the ability of large groups of sources within
overdense regions to deliver ionizing photons into large underdense
regions. Simply put, the common view assumes that reionization
occurred in patches a few Mpc in size, and that it ended nearly
simultaneously in all of them. In reality, however, these two
statements are contradictory. If the patches are a few Mpc in size,
then there is a very large spread in their reionization
redshifts. Conversely, if the spread is small, this implies (from
Figure~3) that the patches must be far larger than is commonly
assumed.

Note that the recombination rate is higher in overdense regions
because of their higher gas density. These regions still reionize
first, though, despite the need to overcome the higher recombination
rate, since the number of ionizing sources in these regions is
increased even more strongly as a result of the dramatic amplification
of large-scale modes discussed earlier.

\subsection{Limitations of current simulations} 

The shortcomings of current simulations do not amount simply to a
shift of $\sim 10\%$ in redshift and the elimination of scatter, for
several reasons. First, the effect that we have identified can be
expressed in terms of a shift in redshift only within the context of
the extended Press-Schechter model, and only if the total mass
fraction in galaxies is considered and not its distribution as a
function of galaxy mass. The halo mass distribution should still have
the wrong shape, resulting from the fact that $\Delta z$ in eq.~2
depends on $M_{\rm min}$. Furthermore, in our more accurate hybrid
model (\S~2.2), the effect on the collapse fraction is no longer
exactly equivalent to a shift in redshift. In any case, a
self-contained numerical simulation cannot rely on approximate models
and must directly evolve a very large volume\footnote{\citet{ciardi}
simulated a 30 Mpc box, larger than previous simulations, but only
gravity was directly simulated, and the mass resolution was three
orders of magnitude lower than the minimum necessary to resolve most
galactic halos at high redshift.}.

The second reason that current simulations are limited is that at high
redshift, when galaxies are still rare, the abundance of galaxies
grows rapidly towards lower redshift. Therefore, a $\sim 10\%$
relative error in redshift implies that at any given redshift around
$z \sim 10$--20, the simulation predicts a halo mass function that can
be off by an order of magnitude for halos that host galaxies (see
Figures~1 and \ref{fig-Lars}). This large underestimate suggests that
the first generation of galaxies formed significantly earlier than
indicated by recent simulations. This makes it easier to explain
recent observations of the cosmic microwave background \citep{WMAP}
that suggest an early reionization at $z \sim 15$--20.

The third reason for the failure of simulations arises from the large
cosmic scatter. This scatter can fundamentally change the character of
any observable process or feedback mechanism that depends on a
radiation background. Simulations in periodic boxes eliminate any
large-scale scatter by assuming that the simulated volume is
surrounded by identical periodic copies of itself. In the case of
reionization, for instance, current simulations neglect the collective
effects described above, whereby groups of sources in overdense
regions may influence large surrounding underdense regions. In the
case of the formation of the first stars due to molecular hydrogen
cooling, the effect of the soft ultraviolet radiation from these
stars, which tends to dissociate the molecular hydrogen around them
\citep{hrl97, rgs02, oh03}, must be reassessed with cosmic scatter
included.

\subsection{Observational consequences} 

The spatial fluctuations that we have calculated fundamentally affect
current and future observations that probe reionization or the galaxy
population at high redshift. For example, there are a large number of
programs searching for galaxies at the highest accessible redshifts
(6.5 and beyond) using their strong Ly$\alpha$ emission \citep{h02,
r03, m03, k03}. These programs have previously been justified as a
search for the reionization redshift, since the intrinsic emission
should be absorbed more strongly by the surrounding IGM if this medium
is neutral. For any particular source, it will be hard to clearly
recognize this enhanced absorption because of uncertainties regarding
the properties of the source and its radiative and gravitational
effects on its surroundings \citep{nature, GRBquasar, s03}. However,
if the luminosity function of galaxies that emit Ly$\alpha$ can be
observed, then faint sources, which do not significantly affect their
environment, should be very strongly absorbed in the era before
reionization. Reionization can then be detected statistically through
the sudden jump in the number of faint sources \citep{hs99,
hai02}. Our results alter the expectation for such
observations. Indeed, no sharp ``reionization redshift'' is
expected. Instead, a Ly$\alpha$ luminosity function assembled from a
large area of the sky will average over the cosmic scatter of $\Delta
z \sim 1$--2 between different regions, resulting in a smooth
evolution of the luminosity function over this redshift range. In
addition, such a survey may be biased to give a relatively high
redshift, since only the most massive galaxies can be detected, and as
we have shown, these galaxies will be concentrated in overdense
regions that will also get reionized relatively early.

The distribution of ionized patches during reionization will likely be
probed by future observations, including small-scale anisotropies of
the cosmic microwave background photons that are rescattered by the
ionized patches \citep{a96, gh98, san03}, and observations of 21 cm
emission by the spin-flip transition of the hydrogen in neutral
regions \citep{t00, cgo02, fsh03}. Previous analytical and numerical
estimates of these signals have not included the collective effects
discussed above, in which rare groups of massive galaxies may reionize
large surrounding areas. These photon transfers will likely smooth out
the signal even on scales significantly larger than the typical size
of an \ion{H}{2} bubble due to an individual galaxy. Therefore, even
the characteristic angular scales that are expected to show
correlations in such observations must be reassessed.

The cosmic scatter also affects observations in the present-day
universe that depend on the history of reionization. For instance,
photoionization heating suppresses the formation of dwarf galaxies
after reionization, suggesting that the smallest galaxies seen today
may have formed prior to reionization \citep{bkw01, s02, b02}. Under
the popular view that assumed a sharp end to reionization, it was
expected that denser regions would have formed more galaxies by the
time of reionization, possibly explaining the larger relative
abundance of dwarf galaxies observed in galaxy clusters compared to
lower-density regions such as our Local Group of galaxies \citep{t02,
b03}. Our results undercut the basic assumption of this argument and
suggest a different explanation altogether. Reionization occurs
roughly when the number of ionizing photons produced starts to exceed
the number of hydrogen atoms in the surrounding IGM. If the processes
of star formation and the production of ionizing photons are equally
efficient within galaxies that lie in different regions, then
reionization in each region will occur when the collapse fraction
reaches the same critical value, even though this will occur at
different times in different regions. Since the galaxies responsible
for reionization have the same masses as present-day dwarf galaxies,
this estimate argues for a roughly equal abundance of dwarf galaxies
in all environments today. This simple picture is, however, modified
by several additional effects. First, the recombination rate is higher
in overdense regions at any given time, as discussed
above. Furthermore, reionization in such regions is accomplished at an
earlier time when the recombination rate was higher even at the mean
cosmic density; therefore, more ionizing photons must be produced in
order to compensate for the enhanced recombination rate. These two
effects combine to make overdense regions reionize at a higher value
of $F_{\rm col}$ than underdense regions. In addition, the overdense
regions, which reionize first, subsequently send their extra ionizing
photons into the surrounding underdense regions, causing the latter to
reionize at an even lower $F_{\rm col}$. Thus, a higher abundance of
dwarf galaxies today is indeed expected in the overdense regions.

The same basic effect may be even more critical for understanding the
properties of large-scale voids, 10--30 Mpc regions in the present-day
universe with an average mass density that is well below the cosmic
mean. In order to predict their properties, the first step is to
consider the abundance of dark matter halos within them. Numerical
simulations show that voids contain a lower relative abundance of rare
halos \citep{co00, som01, mw02, bh03}, as expected from the raising of
the collapse threshold for halos within a void. On the other hand,
simulations show that voids actually place a larger fraction of their
dark matter content in dwarf halos of mass below $10^{10} M_{\odot}$
\citep{gottl03}. This can be understood within the extended
Press-Schechter model. At the present time, a typical region in the
universe fills halos of mass $10^{12} M_{\odot}$ and higher with most
of the dark matter, and very little is left over for isolated dwarf
halos. Although a large number of dwarf halos may have formed at early
times in such a region, the vast majority later merged with other
halos, and by the present time they survive only as substructure
inside much larger halos. In a void, on the other hand, large halos
are rare even today, implying that most of the dwarf halos that formed
early within a void can remain as isolated dwarf halos till the
present. Thus, most isolated dwarf dark matter halos in the present
universe should be found within large-scale voids \citep{infall}.

However, voids are observed to be rather deficient in dwarf galaxies
as well as in larger galaxies on the scale of the Milky Way
\citep[e.g.,][]{BigVoid, eder, gg99, gg00, el-ad, pVoids}. A deficit 
of large galaxies is naturally expected, since the total mass density
in the void is unusually low, and the fraction of this already low
density that assembles in large halos is further reduced relative to
higher-density regions. The absence of dwarf galaxies is harder to
understand, given the higher relative abundance expected for their
host dark matter halos. The standard model for galaxy formation may be
consistent with the observations if some of the dwarf halos are dark
and do not host stars. Large numbers of dark dwarf halos may be
produced by the effect of reionization in suppressing the infall of
gas into these halos. Indeed, exactly the same factors considered
above, in the discussion of dwarf galaxies in clusters compared to
those in small groups, apply also to voids. Thus, the voids should
reionize last, but since they are most strongly affected by ionizing
photons from their surroundings (which have a higher density than the
voids themselves), the voids should reionize when the abundance of
galaxies within them is relatively low. A quantitative analysis of how
the reionization redshift varies with environment may help establish a
common framework for explaining the observed properties of dwarf
galaxies in environments ranging from clusters to voids.

\section{Conclusions}

We have shown that the important milestones of high-redshift galaxy
formation, such as the formation of the first stars and the completion
of reionization, occurred at significantly different times in
different regions of the universe. This conclusion results from the
fact that the temperature threshold, above which cooling and
fragmentation of gas are possible, selects out dark matter halos that
become exceptionally rare at high redshifts. Consequently, density
fluctuations on large scales modulate the threshold for the collapse
of high density peaks on small scales in the exponential tail of the
Gaussian random field of density fluctuations, and introduce a
remarkably large scatter in the abundance of star-forming galaxies at
early cosmic times.

We have developed an improved method to calculate the cosmic scatter
(see \S 2.2). This yields the first self-consistent analytic model
that matches the halo mass function measured in various regions in
numerical simulations that covered a wide range of the parameter space
of region size, mean density, and redshift.

Since the characteristic distance between nearby sources of ionizing
radiation varies widely across the universe, the overlap of the
\ion{H}{2} regions produced by these individual sources in the IGM
occurs at significantly different times in different cosmic
environments. Quantitatively, we find that the spread in the redshift
of reionization should be at least an order of magnitude larger than
previous expectations that argued for a sharp end to reionization (see
\S~3.1).

Current numerical simulations that treat gravity and hydrodynamics
\citep{g00,abn02,yoshida} largely eliminate this real cosmic scatter,
and are artificially biased toward late galaxy formation since they
exclude large-scale modes (see Figures~3 and 4). We find that galaxy
formation within state-of-the-art simulations with $324^3$ particles
is artificially biased to occur too late by a redshift interval
$\Delta z \sim 0.5$ at $z=7$ and $\Delta z \sim 2.5$ at $z=20$. The
box length used in state-of-the-art simulations of reionization
\citep{g00,yoshida} is 1.5--2 orders of magnitude below the minimum
size necessary to treat the scatter reliably, and so alternative
computational schemes \citep{inprep} must be implemented in order to
quantify the implications of the large cosmic scatter on the
reionization history. This scatter should affect the statistical
fluctuations in the number and clustering properties of sources in
surveys with a narrow field of view (such as the Hubble Deep Field),
the luminosity function of Ly$\alpha$-emitting galaxies around the
reionization redshift, the fluctuations in the 21 cm flux produced by
the neutral IGM, the power spectrum of the secondary anisotropies in
the cosmic microwave background, and the present abundance of dwarf
galaxies in various environments (see \S~3.3). Simulations limited to
a small box may be able to study the scatter in the number density of
galaxies by varying the mean density of the box, but such simulations
cannot probe the global structure of reionization since this would
involve the radiative transfer of ionizing photons over distances
larger than the box size.

\acknowledgments

We thank Paul Steinhardt for suggesting to apply our work to galaxy
formation in voids. We acknowledge support by NSF grant AST-0204514
and NATO grant PST.CLG.979414. R.B. is grateful for the kind
hospitality of the Harvard-Smithsonian CfA and the Institute for
Advanced Study, and the support of an Alon Fellowship at Tel Aviv
University and of Israel Science Foundation grant 28/02/01.
A.L. acknowledges sabbatical support from the John Simon Guggenheim
Memorial Fellowship. This work was also supported in part by NSF grant
AST-0071019 and NASA grant NAG 5-13292 (for A.L.).

\end{document}